\documentclass{article}

\usepackage{PRIMEarxiv}

\usepackage[natbibapa]{apacite}
\usepackage[hidelinks]{hyperref}
\usepackage{url}
\usepackage{graphicx}%
\usepackage{multirow}%
\usepackage{amsmath,amssymb,amsfonts}%
\usepackage{amsthm}%
\usepackage{mathrsfs}%
\usepackage[title]{appendix}%
\usepackage{xcolor}%
\usepackage{textcomp}%
\usepackage{manyfoot}%
\usepackage{booktabs}%
\usepackage{algorithm}%
\usepackage{algorithmicx}%
\usepackage{algpseudocode}%
\usepackage{listings}%
\usepackage{tabularx} 
\usepackage{fancyhdr}       

\usepackage{microtype}

\usepackage{enumitem}

\begin{document}

\title{A University Framework for the Responsible use of Generative AI in Research
\thanks{\textit{\underline{Citation}}: 
\textbf{Smith, Tate, Freeman, Walsh, Ballsun-Stanton, Hooper, Lane. A University Framework for the Responsible use of Generative AI in Research. DOI: forthcoming}
}}

\author{
Shannon Smith\textsuperscript{1} \And
Melissa Tate\textsuperscript{2} \And
Keri Freeman\textsuperscript{3} \And
Anne Walsh\textsuperscript{2} \And
Brian Ballsun-Stanton\textsuperscript{4,*} \And
Mark Hooper\textsuperscript{2} \And
Murray Lane\textsuperscript{3} \And
\\
\textsuperscript{1} Research Services, Macquarie University, Sydney
\\
\textsuperscript{2} Office of Research Ethics and Integrity, Queensland University of Technology, Brisbane
\\
\textsuperscript{3} Graduate Research Education and Development, Queensland University of Technology, Brisbane
\\
\textsuperscript{4} Faculty of Arts, Macquarie University, Sydney
\\
\bigskip
Author order was randomised by ChatGPT with the consent of all authors \\
* brian.ballsun-stanton@mq.edu.au
}

\maketitle

\begin{abstract}
Generative Artificial Intelligence (generative AI) poses both opportunities and risks for the integrity of research. Universities must guide researchers in using generative AI responsibly, and in navigating a complex regulatory landscape subject to rapid change. By drawing on the experiences of two Australian universities, we propose a framework to help institutions promote and facilitate the responsible use of generative AI. We provide guidance to help distil the diverse regulatory environment into a principles-based position statement. Further, we explain how a position statement can then serve as a foundation for initiatives in training, communications, infrastructure, and process change. Despite the growing body of literature about AI's impact on academic integrity for undergraduate students, there has been comparatively little attention on the impacts of generative AI for research integrity, and the vital role of institutions in helping to address those challenges. This paper underscores the urgency for research institutions to take action in this area and suggests a practical and adaptable framework for so doing.
\end{abstract}

\keywords{Large Language Models \and generative AI \and artificial intelligence \and ethics \and framework \and responsible use \and research \and research integrity \and university \and research institution}

\section{Introduction}\label{sec1}

Generative Artificial Intelligence (generative AI) poses both opportunities and risks for the integrity and quality of research. To support researchers in using AI responsibly, institutions must adapt their research governance and management practices to mitigate the risks of generative AI while remaining flexible to allow for its innovative use in research. This is a significant challenge due to the high degree of uncertainty of how generative AI will ultimately impact the research sector. This uncertainty is driven by the complex research regulatory environment, the rapid pace of disruption and innovation from AI, and the differing norms between research disciplines.

This paper draws on insights from two Australian universities to provide a strategic framework to navigate this complex and evolving space and to promote the responsible use of generative AI in research. We provide worked examples of how the two universities have followed this framework to develop a principles-based position statement and guide the delivery of communications, infrastructure, process improvements, and training to support researchers in the responsible use of generative AI. Our framework serves as a foundational and adaptable model, offering clarity and direction for research institutions.

The rapid emergence of readily accessible generative AI presents clear opportunities for research analysis, writing, editing, coding, and media creation, which also necessitates robust and ethically sound guidelines for its use. However, although there has been more than 60 years of research on AI \citep{pask_machines_1963}, in a broader sense and in the context of higher education, most recent studies have focused on the impact of generative AI on undergraduate assessment \citep{sullivan_chatgpt_2023}. A systematic review by \citet{crompton_artificial_2023} noted a sharp increase in research on AI, but with a continued focus on the use of AI for undergraduate teaching, learning, and assessment. 

There is also a growing body of literature from stakeholders on the responsible use of AI, for example by funders and publishers \citep[see][]{ganjavi_publishers_2024,teqsa_artificial_2023}, but discussions of institutional responses to the responsible use of generative AI for \textit{research} are much less common \citep{popenici_critique_2023}. This is a gap that needs to be addressed. Universities and research institutions are uniquely placed to instigate proactive measures within the research community to mitigate serious risks posed by generative AI. These risks include potentially fuelling the so-called reproducibility crisis in science, and exacerbating problems such as paper mills \citep{gibney_could_2022, liverpool_ai_2023}. Arguably, these risks to the integrity of research, which threaten its credibility and undermine public trust and confidence in scholarship, are equal in severity to the risks concerning student education and assessment, which are receiving the larger share of attention. 

The aim of this paper, therefore, is to provide structured guidance in approaching this evolving context. The literature and discussion in this paper should help inform institutional decision-making related to governance and policy implementation. We offer a strategic framework that identifies critical policies and practices that institutions and relevant stakeholders must address as a starting point to guide appropriate and effective responses to the opportunities and challenges related to the responsible use of generative AI in research.

\section{Scope}\label{sec2}

This paper presents a strategic framework to help inform university policy and practice in the responsible use of generative AI in research. Each component of our framework is based on current literature and on the experiences of two Australian institutions that have used it to guide and promote institutional research integrity and the responsible use of generative AI in research. We describe the framework in terms of both: (i) the \emph{elements} institutions should consider in fostering the responsible use of generative AI in research, and (ii) the \emph{structure} of those elements within a practical implementation plan considering how the elements interrelate.

Although the examples in this paper describe two specific cases, and both in an Australian context, we argue that the structure of our framework is more broadly applicable, and indeed could potentially be adapted by most large research institutions to meet their own contexts and strategic objectives. A critical element of the framework is the development of an institutional position statement on the responsible use of generative AI in research that applies the principles of research integrity to the specific challenges posed by generative AI within the university context. We explain how a position statement can then serve as \textit{common ground} from which to mobilise training, communication, and infrastructure initiatives to meet specific institutional and research needs. We provide our two position statements as examples.

The scope of the framework we present in this paper is limited to the \emph{use} of generative AI by researchers, including research students, in the conduct of research. We do not consider the ethical \emph{development} of generative AI and its use in emerging technologies, nor do we consider generative AI in the context of research or student services, or its use in university coursework and education, such as in undergraduate assessment activities. 
\section{Strategic Framework}

Our framework is structured in four layers, to be read from the top down to represent the order in which the elements should generally be considered and implemented.

\begin{itemize}
\item
  \textbf{Context} describes the external and internal policy environment that governs research integrity, research conduct, and research student supervision and assessment. Such details help shape institutional responses to opportunities and risks posed by generative AI in research.
\item
  \textbf{Developing a university position} describes the importance of developing a position statement to apply the principles of research integrity to the specific opportunities and challenges posed by generative AI. It describes worked examples by our two universities and how we consulted across disciplines and engaged AI researchers to develop a position statement. The exploration of commonalities and differences between our approaches should be useful for others developing their own position statements.

\item
  \textbf{Implementation} describes a plan to put the position statement into practice with appropriate support, processes and infrastructure.
\item
  \textbf{Review} describes a plan to iteratively evaluate the framework to test its effectiveness and undertake revisions or updates to ensure currency.
\end{itemize}

\begin{figure}[ht]
\centering
\includegraphics[width=\textwidth]{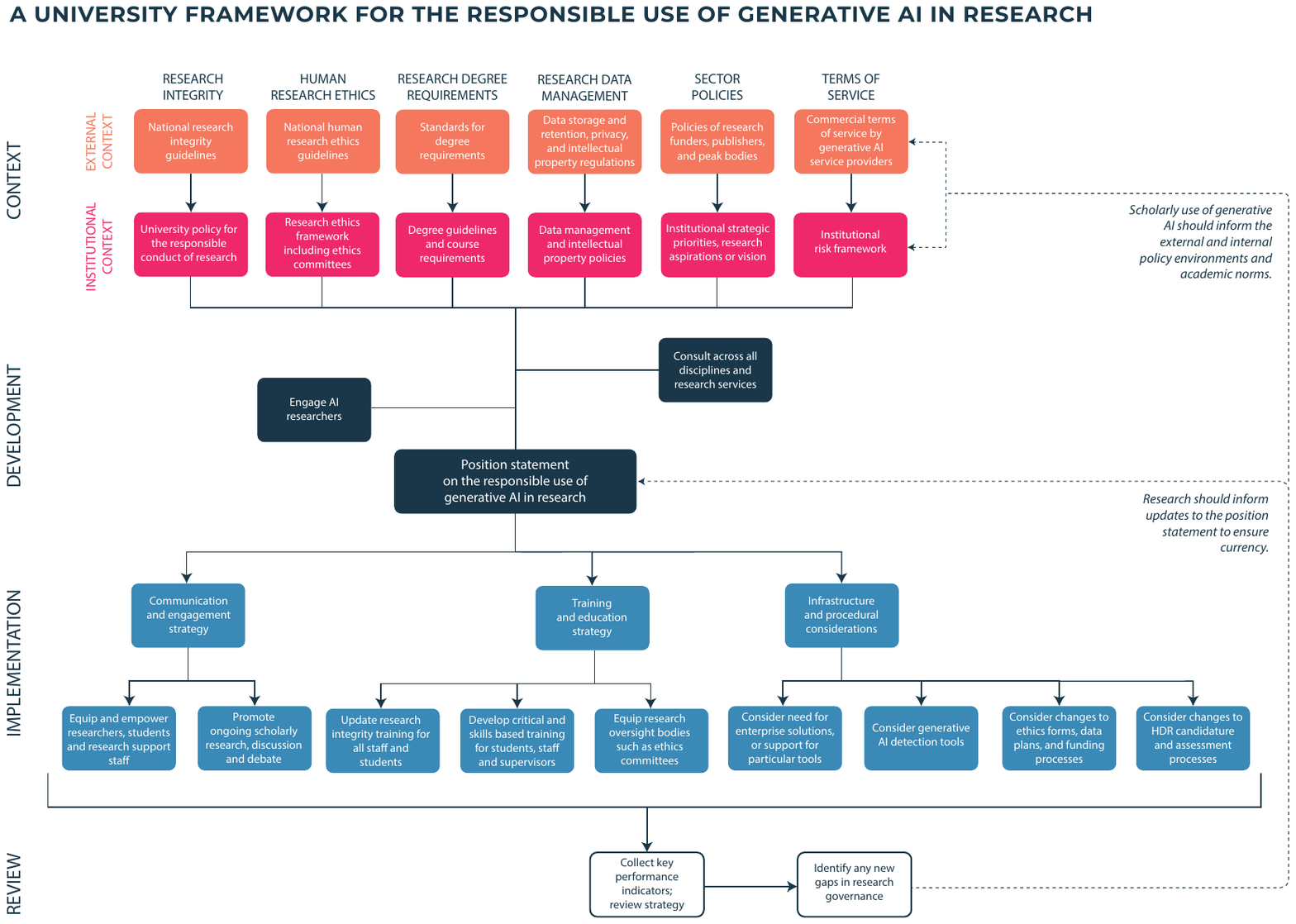}
\caption{A framework for institutions to support the responsible use of generative AI in research. We have made this available to use under a Creative Commons CC-BY 4.0 Licence, accessible via \url{https://osf.io/9b5an/}.}\label{fig:framework}
\end{figure}

Figure \ref{fig:framework} provides a diagrammatic representation of our framework. 

In this section, we describe each layer of the framework in turn.

\section{Context}\label{sec3}

Policy development by research institutions is shaped by external and internal contextual environments, and in the case of generative AI, the competing relationships between these environments are complex. The increasing use of generative AI is creating new ways of conducting research; introducing new techniques, methods and complex datasets. This poses challenges to accepted research practices, and requires institutions to reassess governance frameworks and infrastructure to support researchers and university staff in research integrity, research ethics, data management, intellectual property (IP), copyright, information management, scholarly communication and research degree requirements. 

\subsection{Research integrity}\label{research-ethics-and-integrity}

Globally, various external codes of practice govern institutional research integrity, or what is sometimes called the responsible conduct of research (see Table \ref{tab:tab1}). These codes typically list a number of principles that articulate good research practice or good researcher conduct, and go on to articulate how those principles translate into ethical responsibilities for institutions and their researchers. \textbf{Table \ref{tab:tab1}} lists the  principles of selected international codes, and draws attention to commonalities between them.

In our context, the Australian Code for the Responsible Conduct of Research \citep{nhmrc_australian_2018} establishes eight principles, with which Australian institutions and researchers are expected to comply, irrespective of the technological approaches and tools utilised in the research process. Therefore, Australian research institutions have a responsibility to consider how these broad principles should be applied to govern research associated with the use of generative AI. For example:
\begin{itemize}
\item
  How can researchers apply the principle of \textit{transparency} in disclosing their use of generative AI tools?
\item
  How can researchers take steps to maintain their \textit{accountability} when relying on data generated or analysed by large models?
\item What steps should researchers take to ensure \textit{rigour} in research processes that use generative AI?  
\end{itemize}

Similarly, where a code, council, or funding body describes a responsibility, such as the basic requirement to comply with relevant laws, regulations, and disciplinary standards \citep{nhmrc_australian_2018}, the task of an institution is then to guide its researchers in understanding and considering \emph{which} external laws, regulations, and disciplinary standards might apply to the use of generative AI in research, whilst also harnessing the potential innovative uses of generative AI in research. Effective, ethical, and responsible research demands explicit research integrity guardrails for the use of generative AI in research, and clearly institutional governance, policies, and training, must also reflect the external factors and global concerns related to upholding relevant ethical codes of conduct and principles for the responsible conduct of research. This is a complex task, given the changing dynamics of generative AI across a range of institutional research contexts, and further highlights the need for a strategic framework to address this challenge.

\begin{table}[ht]
\caption{Codes for the responsible conduct of research.}    
\begin{tabularx}{0.95\textwidth} { 
  | >{\raggedright\arraybackslash}X 
  | >{\raggedright\arraybackslash}X |}
 \hline
 \textbf{Document Name} & \textbf{Principles} \\
 \hline
 Australian Code for the Responsible Conduct of Research \citep{nhmrc_australian_2018} & Honesty, Rigour, Transparency, Fairness, Respect, Recognition, Accountability, Promotion \\
\hline
Singapore Statement on Research Integrity \citep{resnik_singapore_2011} & Honesty, Accountability, Professional courtesy and fairness, Good Stewardship \\
\hline
European Code of Conduct for Research Integrity \citep{allea_european_2023} & Reliability, Honesty, Respect, Accountability \\
\hline
UK Research and Innovation Concordat to Support Research Integrity \citep{universities_uk_concordat_2019}& Honesty, Rigour, Transparency and open communication, Care and respect, Accountability \\
\hline
Netherlands Code of Conduct for Research Integrity \citep{knaw_nederlandse_2018} & Honesty, Scrupulousness, Transparency, Independence, Responsibility \\
\hline
French charter for Research Integrity \citep{high_council_for_evaluation_of_research_and_higher_education_french_2019} & Compliance, Reliability of research work, Communication, Responsibility in collective work, Impartiality and independence in assessment and expertise, Collaborative work and plurality of activities, Training  \\
\hline
WHO Code of Conduct for Responsible Research \citep{world_health_organisation_code_2017} & Integrity, Accountability, Independence and impartiality, Respect for persons and communities, Professional commitment \\
\hline
APEC Guiding Principles for Research Integrity \citep{barr_apec_2022} & Honesty, Responsibility, Rigour, Transparency, Respect, Fairness, Diversity \\
\hline
\end{tabularx}
\label{tab:tab1}
\end{table}

\subsection{Trust in Research}

Whilst there is much that an institution can do to protect research integrity and preserve trust in research being conducted by its own researchers, there is wider concern about the overarching influence of generative AI on research trust globally. The democratisation of generative AI and its potential to be used irresponsibly, creates a significant risk to the broad landscape of scholarly publishing. One clear example of this information pollution is the prevalence of paper mills. 

Paper mills involve businesses or individuals that generate or sell fake papers designed to mimic authentic research publications \citep{cope__stm_paper_2022}. The global research community has been increasingly grappling with how to address the issue of paper mills and the apparent increase in fraudulent research \citep{van_noorden_how_2023, cabanac_tortured_2021}. The availability of generative AI is widely recognised to exacerbate this problem \citep{liverpool_ai_2023, van_noorden_ai_2023}. Institutions can play a key role in the global response to paper mills by ensuring graduates and researchers are equipped with the knowledge and skills to critically evaluate scholarly publications to identify characteristics of fake or fraudulent research. Institutions can also support researchers to make raw data available along with their research being submitted for publication, both to comply with new funding and journal requirements and to increase trust and reproducibility in academic outputs. This can assist publishers and editors in making informed decisions about the quality and authenticity of their research. Lastly, whilst the misuse of generative AI can exacerbate the problem of fabricated research, in the future it might also support processes to detect irresponsible research \citep{ben_jabeur_artificial_2023}. Therefore, institutions must not only promote research integrity in the conduct of their own researchers, but more broadly equip researchers to understand and critically evaluate the implications of generative AI for research integrity in the global research sector.

\subsection{Human research ethics}\label{human-research}

Research involving human participants is governed by laws, regulations, and guidelines that are often region or country-specific. In Australia, human research ethics is guided primarily by the \emph{National Statement on Ethical Conduct in Human Research} \citep{nhmrc_australian_2018} (National Statement) which sets out the ethical aspects of the design, review, and conduct of human research. When generative AI is used in research with human participation or data, the ethical implications may require close consideration by Human Research Ethics Committees (HRECs), which in some other regions are known as Institutional Review Boards (IRBs). \citep[][]{zhou_ai_2023,grinbaum_dual_2024,porsdam_mann_generative_2023, herington_ethical_2023} Australian Universities must ensure that they have frameworks and training to equip researchers to apply the principles of the National Statement to any use of generative AI, and to equip HRECs with the knowledge required to review such projects.

One relevant ethical issue that falls under the scope of the National Statement relates to bias and inequality. For example, Obermeyer et. al. \citep{obermeyer_dissecting_2019} revealed bias in an AI tool used to identify patients for follow-up care that reduced the number of Black patients identified for extra care by more than half. Images generated by generative AI are also notoriously prone to bias due to their training data, system prompts, or responses to prior scandals \citep{luccioni_stable_2023, ananya_ai_2024, robertson_google_2024}. Moreover, as these systems are not numerate, any images they produce may conform to the prompt by theme and essence rather than with accuracy, proportion, and detail. While AI produced images may be appropriate for some copyright-free blog headers, they are not presently accurate enough to be appropriate for academic outputs.

On the other hand, Salah et. al. \citep{salah_may_2023} suggest that generative AI could potentially help avoid human bias in some cases, such as using ChatGPT to automate coding and categorisation in qualitative studies to reduce the risk of coder bias, which increases the reliability of research results. In any case, equity and bias are examples of emerging human ethical considerations with which researchers and HRECs need to grapple, and for which institutions need to provide new guidance and/or training.

There is also the matter of necessary technical expertise. In accordance with the National Statement, HREC membership is diverse, including, for example, community representatives, doctors, lawyers, academics, and ethicists. However, such membership may not necessarily include the technical expertise to ethically evaluate research with generative AI. Research institutions may therefore need to reconsider the core membership of their HRECs and animal ethics committees to ensure sufficient knowledge and experience is available. In either case, research institutions will need to provide resources and training for these ethics bodies to enable the informed ethical review of research applications with generative AI.

\subsection{Research Degree requirements}\label{research-degree-requirements}

Standards for postgraduate degrees are generally established and monitored by regulatory bodies external to the university. These standards encompass factors such as the demonstration of academic and research capabilities required to complete the degree, ethical and research integrity considerations, and research quality determinants such as originality and significance. The criteria are often generalised principles that may not always incorporate a sufficient level of detail to encompass issues such as the use of generative AI. In Australia, the standard for doctoral degrees \citep{australian_qualifications_framework_council_australian_2013} includes the application of knowledge and skills with full responsibility and accountability for personal outputs and sufficient cognitive skills for intellectual independence and critical thinking. Such principles imbue a high degree of ethical responsibility on the students themselves, and highlight the need for guidance and training in the responsible use of generative AI.

The implementation of postgraduate degree policy at an institutional level requires more detailed guidelines than those offered by external regulatory bodies, and university protocols on the ethical use of generative AI may be one component of these guidelines. This can be challenging, since the novel position of postgraduate students, acting as researchers in training, can sometimes be unclear. Postgraduate students fall in the overlap between researchers and students. On the one hand, their work must be subject to assessment and degree standards. On the other hand, they are researchers, to be treated as academics and even equals, \citep{tanggaard_survival_2017} which means that if research staff are seizing cutting-edge opportunities using generative AI, then it is natural for an institution to encourage postgraduate students to do the same. This duality of identity presents a unique challenge for internal institutional policies and requires universities to take action to adequately support research students' understanding of their ethical responsibilities when using generative AI. Ensuring high-quality postgraduate research output is a fundamental institutional responsibility and therefore, external research degree standards must be incorporated into university strategic responses for training postgraduates in the ethical use of generative AI.

\subsubsection{Assessment}

The emergence of generative AI has brought to the forefront the limitations of traditional assessment methods in accurately gauging students' knowledge and capabilities. The increased global prevalence of contract cheating and information sharing among undergraduates over the past decade has already cast doubt on the efficacy of these assessments \citep{ellis_infernal_2018, bretag_contract_2019}. With the advent of generative AI, it has become increasingly difficult to maintain that the established modes of assessment for both undergraduate and postgraduate students remain suitable indicators of learning success when originality is viewed as a key component of that success.

Challenges to the concept of originality are not new; they echo the concerns raised during previous technological innovations, such as the introduction of calculators, the transition to PowerPoint, and the challenges posed by students' use of Wikipedia \citep{monaghan_calculator_2016, stern_great_2023, haslam_wikipedia_2017}. The current shift that generative AI poses for university assessment practices is also reminiscent of the transition from handwriting to typewriters to word processors, which changed the nature and extent of others' involvement in the production of knowledge communicated in a thesis. Despite this history, the evolving and complex notion of what constitutes \textit{one's own work} is rarely encapsulated in higher education policies, and this remains true regarding the implications of generative AI \citep{luo_critical_2024}. 

The unprecedented capabilities of generative AI technologies have the potential to transform the way researchers conduct their work. It has completely altered our relationship with texts and challenges the way assessments and theses have traditionally been used to ensure originality and quality \citep{mollick_setting_2023}. The use of generative AI for planning research design, conducting research, or assisting in the preparation of outputs necessitates a reevaluation of the social and ethical construction of a researchers' own work. As such, internal assessment processes are an essential part of a university's AI framework in terms of ensuring students are empowered with the necessary skills to utilise new technology and are equipped with an in-depth understanding of their responsibilities when using generative AI for any component of assessed work. 

\subsubsection{How can an assessment demonstrate learning success?}
\label{learning-success}

In Australia, universities have previously viewed the unauthorised student use of generative AI in assessment as academic misconduct, following nationally accepted mandates associating it with contract cheating. It was legislated nationally in 2020 (TEQSA ACT 2011 (Cth) s5 amendments 2020)\nocite{education_tertiary_2011} that the use of text-based generative AI tools fell within the definition of an \textit{academic cheating service}. This legislation was a strong directive that using such tools was fraudulent and dishonest. Such a view, of foundation models merely being cheaper cheaters, was spurred by tools like paraphrasing bots \citep{birks_managing_2020}. However, this view is now challenged by the increasingly diverse ways in which the tools could be utilised for genuine assistance, in tasks such as searching, referencing, ideation, and editing \citep{grootendorst_bertopic_2022,narayan_can_2022,bommasani_opportunities_2022}. This revised perspective has been further highlighted in Australia, where the higher education governing agency TEQSA \citep{teqsa_artificial_2023} has stated that bans on generative AI are fruitless and universities must reform assessment to prioritise students' AI literacy and ethical decision-making capabilities. 

Any effective policy for assessing postgraduate students' work should be enforceable, and so universities must address the feasibility of detection \citep{lodge_assessment_2023}. However, quantitative measures of the use of generative AI are currently unreliable. Students accused of using generative AI can simply claim that they did not, with no recourse to the educator, and worse, effective users are seldom caught at all. \citep[see][]{liang_gpt_2023, openai_how_2023, sadasivan_can_2023}. Therefore, the criteria and thresholds for the assessment of research degrees, and the way we ensure authenticity of content and student engagement, must match the digital skills and capabilities expected of research students. 

In Australia, we will see a movement towards in-person, verbal defences of student and research work for many reasons, from authenticity to AI-proofing \citep{bending_safeguarding_2023, pearce_rethinking_2023}. These defences, similar to the \textit{thesis defence} seen in American universities and the older traditions of \textit{viva voce} (oral exams/lit. living voice) from antiquity. As  \citet{vinge_rainbows_2006} 
predicted, "[T]he beginning of trust has to be an in-person contact." For universities to acknowledge student capabilities, we will need these in-person milestones to validate a student's intrinsic knowledge and subject mastery outside the context of any tool-assisted capabilities.

Businesses expect graduates to possess the skills to discern when to use and how to critically evaluate the outputs of generative AI, both for employee productivity and to keep pace with the accelerating demands of the industry. In academia, researchers who master the appropriate use of generative AI may outperform their peers, just as knowledge workers using ChatGPT have demonstrated superior quality and efficiency compared to their colleagues \citep{dellacqua_navigating_2023}. 

The unique challenges that generative AI poses for postgraduate assessment therefore require a unified approach to assessment, training, and policy. Given the unique position of postgraduate students as both researchers and assessable students, research institutions must strike a balance between effective mechanisms for fair and legitimate assessment, and ensuring that research postgraduates are equipped with the innovative AI literacy skills required to succeed in their chosen field.

\subsection{Research data management}\label{research-data-integrity}

Generative AI platforms collect user data, including potentially sensitive research data. Depending on the security supporting the generative AI platform, data may be at risk of privacy or copyright breaches due to user error or cyber threats. Furthermore, entering data and information through prompts or other methods may compromise the user's data, or ownership of consequent innovations, as the legal and moral aspects of copyright related to generative AI are presently unclear and inadequately regulated \citep{wipo_ip_and_frontier_technologies_division_generative_2024}. Under the Australian Code for the Responsible Conduct of Research, institutions must provide supporting infrastructure for research data, including guidance and facilities, to safeguard data ownership, security, and confidentiality \citep[Section 2]{nhmrc_management_2019}. Institutions must therefore exercise prudence when employing generative AI to support research data, particularly where the generative AI platform may not apply adequate data privacy, security, and IP standards. 

Typically, an institution can negotiate specific terms of service provisions as part of their purchase of enterprise software associated with research and can reasonably expect that it will stay the same throughout the course of the agreement. However, the current rate of change and the plethora of services means generative AI platforms release frequent updates to their business and personal terms of service. These changes may present a significant legal risk if subsequent updates change IP or privacy provisions for data in ways that compromise security and IP (ownership). Institutions will need to regularly monitor the terms of service of generative AI platforms to maintain control over research data and conform with applicable privacy legislation. As well as robust business practices, this can be supported through institutional frameworks and training that effectively guide the research community on how to protect data and manage risks to research integrity when using specific generative AI platforms. 

\subsubsection{Inputs: Privacy, unauthorised IP access, and Permission}

Prudence should be exercised when employing generative AI in situations where sensitive or confidential business information or intellectual property might be exposed to an AI model through prompts or other methods. Furthermore, when using texts from other authors, it is difficult or impossible to collect informed consent about this use. While there may be claims around \textit{fair use} for published papers, unpublished work, work shared in confidence for peer review, or primary research material may pose specific risks around privacy and the use of this sensitive material within a generative AI model. How can a user, inputting others' research or data ensure that they are acting in accordance with the original intentions or paper's licence?

The main concerns here are fourfold: 1) it is hard for most users to tell, when they use generative AI, what of their inputs will be used for model training, 2) it is challenging to predict where, if anywhere, entered data might emerge for other people, 3) it can be difficult to determine if appropriate security measures and normal bug-prevention techniques have been enacted, and 4) users may not be aware if the risks around any of the prior have been fully communicated to participants to ensure informed consent.

When OpenAI, for example, says ``Opt Out. If you do not want us to use your Content to train our models, you can opt out by following the instructions in this Help Center article.'' \citep{openai_terms_2023} the word ``train'' is being used in a specific technical sense. However, there is not a common understanding about what happens to entered data once it has been ``used'' for training. Companies have not been forthcoming on this point, which increases uncertainty, prevents informed consent, and creates ambiguity for policy-makers. OpenAI's explanation of this process contained no useful operative details at the time of writing \citep{openai_help_center_how_2024}.

Even if we trust that a statistical representation of all text on the internet \citep[][]{chiang_chatgpt_2023} is the default mode of training, and there is no special emphasis placed on \textit{learning from} user inputs, our concerns with the novel threats to privacy may obscure other risks from these systems. Importantly, we may lose sight of the fact that these systems are, at their core, standard server-based systems holding our chats and our data for however long the companies decide to hold them. Users with weak passwords, lack of 2-factor-authentication, or those who share logins with their colleagues are a far more specific and real threat to sensitive data sharing than the quite valid concerns around training. There have also been incidents where these systems, through bugs or poor user security posture, have shared other users' chats without authorisation \citep[see][for an incident this year]{goodin_openai_2024}. University research services need to take these normal threats into account as well.

The fledgling generative AI industry has not communicated its normal processes or risks in a way that researchers can then own the consequences. For example, when an individual, as a private citizen, chooses to use a large language model, they are using it in accordance with their own risk tolerances. However, as research ethics committees are required to protect the researcher, the institution, the participants, and the practice of research, they want to make sure that the researcher fully understands, accepts, and can communicate the risks of their research to any participant or involved party. This is a difficult job when the technologies and their terms of service change regularly. When making a position statement, universities will therefore need to weigh the shifting environment of these technologies, their terms of use, the ability of researchers to own, understand, and communicate risks, and the public norms around data sharing and processing. Each institution will have its own risk appetite and its own demands for efficiencies, but these risks should not be tacitly accepted.

\subsubsection{Outputs: Copyright and IP sharing concerns}

Intellectual property can be legally protected, for example through copyright, trademarks or other IP rights such as database rights or industrial designs \citep{wipo_ip_and_frontier_technologies_division_generative_2024}. The legal aspects of copyright related to the outputs of generative AI in Australia have yet to be updated to address the use of generative AI, and are presently unclear and inadequately regulated \citep[]{samuelson_generative_2023}. Another concern is some companies responsible for creating foundational generative AI models have recently faced legal disputes concerning copyright infringements related to utilisation of materials for AI training. It is possible that liability could extend to the users of the models. It is not currently known whether works created by generative AI platforms will be protected by copyright. Copyright will only subsist in works that have a human author who has contributed 'independent intellectual effort' to the work. It is possible that works generated by AI do not meet the human author threshold and may not be eligible for protection under the current law. This highlights copyright as a critical ethical and legal concern that may create uncertainty for researchers. The need for prudence by researchers, with guidance from their research institution,  is particularly important since copyright regulations seem to be lagging. This is another example of the need for researchers to be supported while clearer legal precedents and specific regulations addressing the use of generative AI in research are addressed and underscores the importance of research institutions providing clear policies, frameworks, and guidance to support researchers in navigating these complex issues and ensuring compliance with ethical and legal standards. By proactively addressing copyright concerns and offering resources to researchers, institutions can foster responsible and innovative use of generative AI while mitigating potential risks.

The challenges currently posed by generative AI platforms in the issue of research data management underscore the significant role of institutional policies, guidance, and training in supporting researchers. By proactively addressing data privacy, security, and IP concerns, providing clear guidelines, and equipping researchers with the necessary knowledge and skills to navigate this evolving landscape, institutions can foster a culture of ethical and responsible research practices. These efforts are essential elements of the overall framework for promoting the responsible use of generative AI in research.

\subsection{Sector policies}\label{sector-policies}

\subsubsection{Publishers}\label{publishers}

Academic publishers are responsible for applying high standards of publication ethics. Journal policies support responsible research practices related to accountability and transparency, and many journals that are members of the Committee on Publication Ethics (COPE) must also adhere to their guidelines \citep{committee_on_publication_ethics_cope_2024}.
Springer Nature and Elsevier revised their author guidelines, and in some instances have stipulated that generative AI cannot be used or credited as authors. However, it may be possible to use generative AI in the writing process to edit grammar and check for readability if this is correctly acknowledged \citep{nature_portfolio_artificial_2023}. 

The Nature portfolio's position on generative AI also highlights accountability as a research integrity principle and confirms that services offering large language models, such as ChatGPT, cannot be held accountable for attribution of authorship \citep{nature_portfolio_artificial_2023}. The publisher requires increased transparency with any use of generative AI to be documented in the Methods section \citep{nature_portfolio_artificial_2023}. However, Nature forbids the use of generative AI in image and video production, due to concerns over copyright and research integrity. The rules around acknowledgement of use are also part of the OpenAI Sharing and Publication policy, which states that humans ``must take ultimate responsibility for the content being published" \citep{openai_sharing_2022} and any use of generative AI for drafting or editing purposes must be detailed in the Foreword or Introduction sections. 

As it is difficult to enforce any policy against the use of generative AI, Nature Journals' and Open AI policies highlight essential and pragmatic practices for researchers' effective and responsible use of generative AI \citep[][as well as our discussion in \S{}\ref{learning-success}]{kirchner_new_2023}. Firstly, that responsibility for the content must always fall upon the person who is claiming authorship. Secondly, the need to acknowledge how generative AI has been used in writing the publication. These practices reinforce the principles of accountability, honesty and transparency that underpin institutional position statements. At the same time, given the rapidly evolving landscape of generative AI, publishers and institutions must be careful to avoid harming authors with false claims of inappropriate use of generative AI \citep[see][]{wolkovich_obviously_2024}.

\subsubsection{Research funding bodies}\label{arc}

Large research funding bodies external to universities are naturally influential on institutional research policies and practices and may dictate national policies for responsible use of research funds. In China and the U.S., policies related to the use of generative AI are being shaped, and to date, it appears the European Union will be one of the first jurisdictions to implement strict legislation to minimise the risks generative AI poses for transparency, bias, privacy, and copyright principles, in particular for generative AI models that generate images and video \citep{gibney_what_2024}. In Australia, many funding bodies appear, so far, not to have taken a general stance about the use of generative AI in the research they fund, but several have implemented policies relating to the use of generative AI in writing and peer reviewing grant applications \citep{australian_research_council_research_policy_branch_policy_2023, national_institute_for_health_use_2023}.
The Australian Research Council (ARC) acknowledges that generative AI may present opportunities for research and for the writing of grant proposals, but maintains that the responsibility for any errors or IP violations falls squarely with the researchers and their institutions. This may not be a new stance regarding responsibility for funds, but in light of the complexity and uncertainty introduced by generative AI, it places an additional burden on institutions to undertake due-diligence, and to provide appropriate training and guidance. 

Although, several funding bodies have prohibited the use of generative AI for peer review, recent controversy has suggested that some reviews have been poorly generated by reviewers using ChatGPT \citep{lu_are_2023}. While generative AI \textit{can} mimic patterns of text which have occurred in judgement-using language, it cannot serve as a reliable peer assessor at the current capabilities of frontier models. Peer review feedback entails a judgement related to the pragmatic realities of the grant, and generative AI cannot effectively make those judgements which require conceptual knowledge. Furthermore, by default, Microsoft Copilot and OpenAI use text entered into their tools for model training, and while it is possible to act with privacy-preserving choices as part of their fee-paying services, the authors of the grants under review may not have given permission for their text to be used for training, and it is unlikely that peer reviewers can grant such a licence. There is also serious concern that research grant applications contain information and intellectual property about innovations that have not yet been published and/or entered into the public domain. We can imagine a world where a reviewer would be able to use these tools to learn more about the details and nuances of the topic under review. We can also imagine a world where computer-aided or computer-replaced review can serve to mitigate some issues with present peer review \citep{horrobin_peer_1996,tennant_limitations_2020, scott_peer_2007}, but these technologies cannot yet substitute for the expert human judgement of peer review, even if the default use of them is privacy-preserving. To do so, would undermine confidence in peer review for assessing grants and the responsible use and allocation of research funds. This corrodes public trust and confidence in research. Again, this highlights the important principle of accountability in institutional position statements for the responsible use of generative AI as it applies to research funding practices.

\subsubsection{Institutional strategies, operational plans, and research 
}
In the Australian context, a 2024 government review of the quality, accessibility, and sustainability of higher education has highlighted the potential of generative AI to increase research productivity and directed Australian universities to take the lead in building new forms of research infrastructure \citep[62]{australian_universities_accord_australian_2024}. Universities articulate their overarching, and often progressive, research aspirations and strategic priorities in institutional strategy documents. To realise these goals, Key Performance Indicators (KPIs) that measure success against these strategies may then be articulated in operating plans. In Australian institutions, KPIs relating to research achievements frequently include measures such as external research income, research publication quality and quantity, graduate research completions, and indicators of world standing, such as global rankings. 

Some of the performance measures listed above are influenced by, and have impacts upon, the use of generative AI in research. The use of generative AI has the potential to support key strategic objectives related to research efficiency; for example, generative AI may potentially enhance the efficiency of the peer review by manuscript screening, language quality assessment, and reviewer selection \citep{calamur_adapting_2024}. Large language models may also have a role to play in digital transformation, digitalisation, and digitisation strategies: a reprise of the attempts for a \textit{paperless office} in the 1990s \citep{meyendorf_digitization_2021}. Most obviously, universities may attempt to deploy chatbots to reduce query load on their professional staff. It is likely that this \textit{efficiency dividend} will also be expected of research integrity and ethics committees. 

In the context of institutional research strategies and aspirations, there is a risk that universities may not accurately assess the importance of human judgement in their research support and governance roles. A large language model might provide the most probable words, but not necessarily the most considered ones. For some research institutions, the fast pace of change in this area will not only necessitate a shift in focus of already in-flight projects associated with areas such as digital transformation and postgraduate student support or assessment, but also cyber-security. The pace at which the industry of generative AI is moving may also pose dangers to the sometimes cumbersome enterprise software purchasing process. For instance, by the time products are evaluated, procured and implemented, the generative AI platforms on which they are based could have changed significantly. The impacts that the rapidly evolving generative AI context will have on institutional research objectives and operational plans require proactive strategies to ensure institutional frameworks incorporate the principles of responsible research conduct and adequately support researchers.  

\subsection{Terms of service}

The terms of service offered by service providers are also an essential, if highly technical factor, when considering a university's position statement. When researchers seek to use large language models, the following factors must be considered: 1) the service provider (e.g. OpenAI, Anthropic, Microsoft, Perplexity.ai) and the terms of service which bind all interactions between the provider and the user, 2) if the service provider is running the large language models on their own hardware or if they are, themselves, bound by third-party terms of service agreements, 3) the version of the large language model and if payment status changes the specific terms of service used (i.e. the public and free version of Microsoft Copilot offers far less privacy and IP preserving terms than the business for-pay version), and 4) the mechanism by which the researcher is interacting with the service.

The final point around the mechanism of interaction requires some elaboration. While these large language models may present to most users as a chatbot style interface, for those with unusual, computationally controlled, or considerable workloads, a computer-to-computer interface called an Application Programming Interface (API) is available. In contrast to the chatbot mode, these APIs tend to be pay-as-you-go. They are also generally far more privacy-preserving than the chatbot mode, even if all other aspects of the service provider and model are the same. 

Researchers, and research data managers will also need to pay close attention to the terms of service over the lifetime of their research projects.  As new models are seemingly released monthly, and service offerings and providers change equally regularly, as previously discussed, the companies involved frequently update their terms of service, and these updates are usually not announced to their customer bases. Thus, what might be a privacy-protecting terms of service when a project begins, may not be by the time the project ends. Position statements should clearly allocate the responsibilities of monitoring the offerings of these generative AI companies and to maintain a list of services, models, and modes of interacting which are risk-acceptable and privacy-acceptable to the university.

AI platforms also mandate service use disclosure that puts the onus of responsibility for any text generated by large language models onto the human author \citep{openai_sharing_2022}. An awareness of the language, terms, and contracts with the enterprise sales teams of these services informs the presented framework and the exact nature of the guardrails it provides. Specifically, research institutions should regularly monitor the terms of service of all approved generative AI platforms. While institutions cannot forbid an entire technology category, an institution can point to specific ways of using specific services with alarm and guide researchers to more effective and less risky options. The continually changing technology and changes in terms of service require that maintaining these position statements is an ongoing task.

\section{Developing a University Position }

Considering the diverse contextual factors relevant to the use of AI in research,  as specified in the strategic framework we offer, a position statement can serve the important function of distilling and applying that context in one place. For an institution with obligations to guide its researchers in responsible conduct and compliance, endorsing a clear position statement, or another policy instrument that serves the same function, is essential. A position statement also provides a foundation for the development of institutional policies, guidelines, and training specifically related to research integrity and the responsible use of AI.

A position statement is an opportunity for an institution to take a proactive stance to promote the responsible use of generative AI insofar as it fosters research quality and innovation, and to prohibit certain practices if they are inconsistent with research integrity. The act of developing a position statement can itself help to educate research managers and research leaders, and empower them to contribute to institutional initiatives. By providing a common ground within the institution, a clear position statement can reduce confusion caused by the saturation of literature and opinions, the proliferation of new tools, and the rapid pace of disruption. It can guide researchers in understanding the relevant policies and ethical considerations that apply to their specific context. 

A position statement can be simple, with clear declarative statements about responsible use, that can then form the basis for more complex and specific considerations and scholarly debates about more nuanced matters. Lastly, a position statement focusing on the use of generative AI in research can help ensure researchers are considered in any whole of enterprise strategies dealing with the governance and risk associated with generative AI. In the following section, we explain how our two institutions each developed position statements of this kind and for these goals, and we provide our position statements as examples.

\subsection{Consult across disciplines and research services}

Developing a position statement requires broad consultation across research disciplines and research support services to capture the diverse ways of using generative AI in research, including targeted consultation with academics who have expertise in research ethics and responsible research practices. This consultation could take multiple forms.  

\begin{enumerate}
    \item The Queensland University of Technology (QUT) began consultation in January 2023 by drafting a discussion paper: \emph{Emerging AI tools and challenges for research integrity}, which was then circulated broadly to provoke discussion and feedback. This discussion paper recognised the potential opportunities of emerging generative AI tools in research, and the inherent risks generative AI tools can pose for responsible research practices such as in authorship, originality of publications, and research outputs. It provided a balanced overview of the use of generative AI tools in research to help focus the consultation on the key issue of how generative AI tools intersect with research integrity. Feedback was welcome from all parts of the university. In particular, targeted feedback was requested from Research Integrity Advisors (RIAs) and Research Ethics Advisors (REAs). These advisors are academics appointed by the university to provide expert advice about responsible research practices within each faculty and school.
\item Macquarie University (Macquarie) established a committee of stakeholders and consulted experts from each faculty including RIAs and strategic professional staff to consider the risks and opportunities posed by generative AI to research integrity, and various research processes and operations. The committee was tasked with considering whether a new policy or targeted policy updates were required, or whether changes to processes were deemed appropriate. Anticipating the fast pace of change and the forecast need for regular updates, a Guidance Note outlining the University's position and recommendations regarding the use of generative AI in research was drafted. To create the Guidance Note, the committee appraised the current literature and engaged in cross-faculty discussion with researchers and research leaders. A key aspect of the Guidance Note was the discussion and identification of key resources to provide an initial foundation for researchers, leaders, and support staff who wanted to learn more. Because each faculty was taking its own approach towards the use, promotion, and experimentation around these tools, the Guidance Note had to provide a summary of the risks and justifications for the intended approach, such that a guiding coalition was developed among its readers. Following standard governance pathways, the Guidance Note was submitted to the relevant committees for noting or endorsement. This step was considered critical to enable diverse areas of the University an opportunity to build understanding and consider implications for their areas or alignment with other projects or initiatives. 

\end{enumerate}

In both cases, consultation was coordinated by each university's Office of Research Ethics and Integrity (OREI), as the institutional office responsible for fostering and promoting research ethics and responsible research practices. The consultation process provided the opportunity for OREI at both universities to engage with their research communities to understand the varied uses of generative AI in research, and the potential impacts of generative AI tools on responsible research practices, and to promote the core values of honesty, transparency and accountability inherent in research practice. In this way, the consultation process reasserted the values of social trust and accountability \citep[Chapter 3]{connors_encouraging_2002}. 

The examples from QUT and Macquarie illustrate that while the approach to developing a university position can vary according to each institution's needs and internal structure, coordinated consultation, discussion, and feedback across university research communities is crucial to ensure all stakeholders are provided the opportunity to participate in the process. By conducting broad consultations and engaging diverse perspectives and expertise, universities can develop a position statement that will promote the responsible use of generative AI in research and foster a shared understanding of the associated opportunities and challenges.

\subsection{Engage AI Researchers}

Apart from the formal consultation process across research disciplines and services, as summarised above, it is vital to recognise that some researchers have long been experts in this space, and so staff may already be engaging both internally and publicly with emerging opportunities and challenges. Of course, wherever possible, these academics should be invited to participate in the consultation process. However, some researchers may be wary of internal policy initiatives, and may therefore be reluctant to participate, especially if there is any perception that there will be a prohibition on generative AI, inconsistent with some researchers' own views to harness the benefits of generative AI for research and innovation. 

Further, some researchers may already be publishing their own guidelines or ethical frameworks. There is a long history of scholarship seeking to define principles for the ethical \emph{development} of AI \citep{dreyfus_mind_1986}. Now that commercial generative AI tools allow such widespread use, scholarship has naturally extended further into ethical frameworks for its \emph{use} \citep{schlagwein_chatgpt_2023, bell_rapid_2023, dwivedi_opinion_2023}. For institutions with scholars in this space, it is important to work with them where possible, and to promote and acknowledge their expertise from an early stage. Irrespective of institutional policy, academic staff and students are much more likely to learn their research norms from fellow academics and research supervisors than from centralised initiatives or training \citep{faden_importance_2002, godecharle_heterogeneity_2014}. 

For any research integrity initiative, there is a risk of disconnect between those promoting responsible principles, and the cohorts being asked to practise those principles in their day-to-day research \citep{kretser_scientific_2019}. This risk can be mitigated not only through consultation and socialisation, but by supporting researchers' own initiatives and presentations, actively promoting their work, giving them a platform, and offering to co-jointly deliver training initiatives with them. Engaging AI researchers within the university can form a vital part of an institution's strategic framework in shaping policies related to research integrity, providing essential AI literacy training for researchers, and ensuring that institutions not only address the challenges, but also explore the opportunities that generative AI may provide.

\subsection{Position Statement}

 A strong position statement should address the key research integrity principles of honesty, transparency, and accountability, and reinforce the importance of these principles in the responsible use of generative AI in research practices. In the context of the strategic framework suggested in this paper, a position statement on the use of generative AI in research is a guideline based on diverse inputs and should provide a consensus among the relevant stakeholders, including academic staff and research students \citep{aaps_open_editors_aaps_2023}. To that end, our institutions' position statements have been developed based on broad stakeholder feedback, and aim to provide both a consolidated institutional position and clear guidance for diverse faculties and research groups. We provide copies of our position statements as Attachments 1 and 2, and make them available to adapt under a Creative Commons 4.0 CC-BY
Licence.

Broadly, our two position statements have much in common:

\begin{itemize}
\item
  They are designed to support policies about the responsible conduct of research. In this sense, they are supplements to, or explanatory notes about, how research integrity policies should be understood in the context of generative AI.
\item
  They \textit{encourage} the use of generative AI in research where it promotes research quality and impact, where it can be used responsibly.
\item
  They are principles-based and are not primarily designed to provide a set of rules, although both statements do mandate some aspects such as:

  \begin{itemize}
  \item
    {}A generative AI tool must not be listed as an author, as this violates the principle of accountability for research outputs.
    \item   
    {}Use of a generative AI tool in the Methods or Acknowledgements sections of a research output are consistent with the policies of publishers. This is consistent with the principle of transparency in the use of AI.{}
  \end{itemize}
\end{itemize}

There are some notable differences between the documents in terms of style and delivery, which can largely be attributed to the different contexts in which they were developed:

\begin{itemize}
\item
  While QUT named its document a ``position statement" to emphasise that it provides definitive university-level advice, Macquarie University named its statement a ``Guidance Note", emphasising that its role is to support a broader policy about the responsible conduct of research and is likely to require updating at regular intervals. To some extent, the position of the University was also outlined in the accompanying paper provided to relevant committees. 
\item
  Macquarie University's Guidance Note recognised that flat and unqualified policy statements would not persuade staff and students in the absence of associated training and support, resources for which could not at the time be allocated. Therefore, Macquarie's note explicitly extends towards educative information about generative AI, whereas QUT has achieved these same goals in a different way: it limited its statement to be a more traditional policy instrument, a summary of responsible use, and provided explanatory and pedagogical support via other means including the discussion paper mentioned previously. 
\end{itemize}

\section{Implementation}

A position statement is only a starting point for effective institutional governance and policies for the responsible use of  generative AI in research. Once established, the university position must be accompanied by initiatives to support the implementation phase within the overall strategic framework, which focuses on training and education, procedural updates, and impacts on infrastructure.

\subsection{Communication and engagement strategy} 

The rapid pace of change in the generative AI landscape, with the frequent release of new services, updates to large language models, and the emergence of models capable of processing increasingly diverse types and sizes of media, can be overwhelming even for those tasked with staying informed. To ensure that researchers can make informed decisions about their time and attention based on a realistic understanding of the current state of the art, the deployment of a position statement must be accompanied by a well-designed communication and engagement strategy.

Introducing a new policy naturally necessitates communication of that policy. Moreover, a policy by itself is insufficient to fundamentally change a culture, so any communication must be accompanied by \textit{engagement} and other initiatives to support adoption. QUT's position statement ``recognises that the use of, and implications of using, generative AI tools may differ widely by discipline, and ...encourages ongoing scholarly debate, discussion and research about these issues." Usually, students and faculty have some choice in how and when to adopt new technology, following the categories of the technology adoption model \citep{singh_22nd_2023,chocarro_teachers_2023}. However, since generative AI poses risks to researchers and institutions, and may be used by students well, poorly, and/or with or without permission, change management is necessary. 

Retraction Watch maintains an ever-growing list of papers with \textit{ChatGPT Writing}, where the search strategy seems to be around identifying indicators of lazy copy-paste practices from these tools \citep{marcus_papers_2024}. The publication of such low-effort output can be deeply embarrassing for researchers and their affiliated institutions. To mitigate these risks, it is pragmatic and necessary for institutions to engage in effective communication with all staff to establish appropriate expectations regarding the capabilities and responsible use of these tools. 

In a classroom context, some professors have attempted to use generative AI to catch students engaging in unauthorised use, often with disastrous results \citep{klee_professor_2023}. These failures can be attributed, in part, to a lack of effective guidance and communication from universities about these tools; such guidance is not necessarily to teach academics the intricacies of appropriate use, but rather to promote a broader understanding of their capabilities and limitations. As new capabilities, services, and models are rapidly becoming publicly available in the generative AI space, institutions must effectively communicate the implications of their policies, deploy large language models to staff and students to mitigate privacy risks and enhance equity, and engage with researchers and departments across the institution to ensure all researchers are aware of the potential risks and benefits associated with the use of generative AI in their research practice.

To achieve effective communication and engagement, we recommend universities plan a coordinated change-management strategy as part of policy development within the overall strategic framework to promote the responsible use of generative AI \citep{nosek_strategy_2019, kotter_leading_1995, moen_circling_2010}. At the very least, for example, authors of the policy should attempt to communicate it \textit{and} the technical or pragmatic reasons for specific clauses as part of department meetings. The policy announcement should also be covered by not only an all-staff communication, but a detailed communication plan that will result in most researchers being exposed to the policy. The publication of the policy can open up opportunities for all-of-university discussions about nuance and implementation, so long as appropriate resources exist to channel that interest into training and further engagement, especially to those critical of the technologies. Through these conversations, a university can refine its understanding of norms accepted by its various faculties and come to a local and nuanced consensus about appropriate and ethical use of these technologies. These cultural norms will likely be more detailed than a policy, and can also serve to mitigate risk, and understand the training required to support researchers' AI literacy development.

\subsection{Training and education strategy}

Training and education are fundamental and ongoing elements of any institution's response to ensuring its researchers fully understand their ethical responsibilities if and when they use generative AI in their research practices. In Australia, institutions have a responsibility to ``\emph{Provide ongoing training and education that promotes and supports responsible research conduct for all researchers}'' \citep{nhmrc_australian_2018}. Moreover, AI literacy is fast becoming an essential part of research processes and a valuable industry skill \citep{ahmed_growing_2023, nordling_how_2023, van_noorden_ai_2023, hutson_hypotheses_2023, prillaman_is_2024}. There is an urgency for institutions to provide education that equips research graduates for their careers and empowers academic staff to pursue industry research partnerships. A 2023 survey of 1,600 researchers \citep{van_noorden_ai_2023} has found that researchers are already seizing the benefits of generative AI for their research, including faster data and computation processing, time and monetary savings, and research breakthroughs where progress was previously infeasible. 

The diversity of opportunities provided by generative AI heightens the importance of AI literacy training and initiatives which emphasise research integrity. Any practical training should be \textit{consistent with}, and more ideally \textit{integrated with}, training about responsible use. In this sense, the research training needs for generative AI are complex. As our framework suggests, it must encompass discipline-specific research processes and publication practices. Currently, there are tensions between acknowledgement of the researchers' use of tools, transparency around shifting norms of use, and the overwhelming need for researchers to not \textit{cite} generative AI produced text as it cannot be an authority. Beyond these AI-specific tensions, the use of these tools also exacerbates problems with current data management policies, authorship guidelines, copyright law and research reproducibility. 

Beyond \textit{knowledge} about these areas, AI literacy training must also focus on the development of \textit{skills} to understand, identify, and address the trustworthiness and reliability of generative AI tools. Necessary researcher skills include the ability to tackle challenges such as privacy and the risk of disclosing IP, the potential for bias and prejudice in the data generative AI tools use to formulate responses, and the ability to critically evaluate outputs. There is a broad range of training required to equip researchers with the generative AI literacy skills to effectively navigate opportunities and risks of incorporating generative AI into their research practices. Therefore, training should focus on explicit scaffolding and self-regulated learning that encourages critical engagement and agency in decision-making when using generative AI tools \citep{markauskaite_rethinking_2022, mcknight_electric_2021, unesco_chatgpt_2023}. Researchers must be equipped to critique outputs, and understand the potential for generative AI to ``hallucinate'' or make up facts. Generative AI does not have a moral compass, as any of its responses are significantly influenced by the tacit and explicit direction of users' prompts \citep{ceron_beyond_2024}.  Its responses can be inaccurate and incomplete \citep{bearman_preparing_2020}. To put the point generally, training must provide mechanisms for researchers to improve their practical generative AI competencies in a complex ethical context \citep{bearman_preparing_2020, gasevic_empowering_2023, unesco_ai_2021, venaruzzo_embracing_2023,feldstein_chatgpt_2023,  mcknight_eight_2022, mollick_using_2023, sullivan_chatgpt_2023}.

Given that the foundations of our universities' position statements are key principles of research integrity; honesty, transparency, accountability and fairness, it is natural that research institutions should also frame core training initiatives to support generative AI in this same way. At our two universities, initial researcher training has included workshops and online modules. Examples include expert panel discussions that have allowed postgraduate students and supervisors to ask questions of panellists with diverse expertise in AI research, research ethics, and student policies. QUT has also produced online modules that allow asynchronous access to a learn-on-demand platform. These modules not only provide an overview of the university position statement, but challenge participants to make decisions for themselves on the appropriate use of generative AI, using the key principles of honesty, transparency, accountability and fairness, through scenarios and case studies. Participants are asked to consider the best action in certain scenarios, and how generative AI could be utilised for their own research.

Generative AI is already influencing how individuals work and conduct research, and it has the potential to transform how institutions operate. A UNESCO report on AI and Education \citep{unesco_ai_2021} has highlighted the existing tension between the promise of generative AI as a tool to enhance, and even transform learning, and the challenges to equitable access and the risks of unethical use. While \citet{unesco_ai_2021}
advised that AI can help achieve its Sustainable Development Goal Four of ensuring \emph{inclusive and equitable quality education for all}, it also acknowledges that AI tools raise profound ethical questions. A strategy advised by Bearman and Ajjawi \citet{bearman_learning_2023} 
to facilitate the effective use of generative AI tools, is not to focus on the idea that everyone will cheat. Instead, training and support should provide quality learning opportunities that facilitate the development of the decision-making skills and strategies required to understand the benefits and limitations of AI tools and to critique and correctly attribute AI outputs \citep{lodge_assessment_2023, teqsa_artificial_2023, unesco_ai_2021, unesco_chatgpt_2023}.

The rapid pace of change and the novelty of generative AI necessitate a comprehensive change management approach that goes beyond the typical scope of university training initiatives. Researchers require tailored training to effectively navigate the unique challenges posed by generative AI, as these technologies introduce unprecedented complexities and ethical considerations. The training strategies themselves must be agile and adaptable to keep pace with the frequent updates and advancements in the field. Moreover, the development of robust policies is equally crucial to guide researchers in the responsible use of generative AI. Failing to address and manage stakeholder needs in this context can have significant consequences, underscoring the importance of proactive and comprehensive institutional responses, and an overarching strategic framework that considers both institutional governance and policy and researchers' AI literacy training.  

\subsection{Infrastructure and procedural considerations}

Another challenge for universities is ensuring equitable access to generative AI tools in a rapidly evolving technological landscape. For this reason, the strategic framework universities implement to guide their response to the ethical use of generative AI in research must address the potential for disparities in knowledge, skills, and access to these tools, as this will lead to significant inequalities between researchers who effectively utilise generative AI and those who do not. Moreover, the swift pace of development and frequent updates to generative AI platforms necessitate institutional policies that are adaptable and responsive to change. To address these challenges, institutions must not only provide guidance on the ethical use of generative AI, but also the infrastructure (access and computational resources) to ensure that all researchers have access to appropriate generative AI tools and the necessary training to use them responsibly. 

AI literacy training will also need to be regularly updated to keep pace with advancements in the field. Furthermore, universities should emphasise that it is in each individual's best interest to use these tools appropriately, as the responsible use of generative AI is not only an institutional expectation but also a personal responsibility that directly impacts the integrity and quality of their research outputs. Institutional policies must be designed with the flexibility to accommodate the rapid evolution of generative AI technologies while still maintaining a strong foundation in the principles of responsible research conduct. A policy, communication strategy, and change-management process are all necessary but not sufficient items for deploying large language models to a university. The use of these tools is also governed by other elements of university policy. For example, at Macquarie:
\begin{quote}
\noindent6.8 Provide access to facilities for the safe and secure storage and
management of research data, records and primary materials and, where
possible and appropriate, allow access and reference.

\noindent and

\noindent 6.9 Facilitate the prevention and detection of potential breaches of the
Macquarie Research Code.
\end{quote} 

Both the aspects above require attention as part of a policy deployment around the responsible use of generative AI. Macquarie would have an obligation to communicate available model choices, trade-offs, and infrastructure available for running or hosting more esoteric models. This infrastructure may be needed to process sensitive or highly sensitive data, or to deal with unusual subject matter. There is also the expectation that as a work-related product, researchers should not have to pay for their own subscriptions. Another obligation, this time in relation to records is the capacity for logging and appropriate sharing of conversations with these chatbots. Because the local norm in 2024 is that archives of all chats must be shared as supplemental data, the university should have some expectation that it will be involved in the hosting and management of that data.

While Microsoft and OpenAI seem to have defaulted to claims of we will not log anything as part of your enterprise subscription \citep{openai_business_2023, microsoft_copilot_2024}, that requirement might not align with the expectations of the Australian Research Code. Therefore, there are infrastructure negotiations which must occur to allow for tool use to be appropriately audited and secured. At Macquarie, we have an obligation to have these conversations as part of our statutory requirements around the prevention and detection of breaches of research integrity. To be clear, in most cases, the poor use of a large language model \textit{is not grounds for findings of academic misconduct.} It is merely poor academic conduct. However, when these breaches may involve the unauthorised or unintentional transmission of sensitive or highly sensitive data, research integrity units have an obligation to investigate exactly what was communicated to the chatbots. Therefore, because institutions need to supply these models; train faculty, staff, and students in these models; capture, secure and share chat logs; and be able to investigate breaches of research misconduct, there are infrastructure obligations which are assumed with most publications of large language model policies at universities. Consideration of infrastructure needs and equitable access to generative AI forms a key part of an institution's overall strategic response to the responsible use of generative AI by its researchers.

\section{Review}

The process of evaluation is an important component in the design, implementation, and refinement of university policy, so is integrated into the framework at critical points. This process provides a review mechanism to inform stakeholders of policy effectiveness, helps to ensure currency of the position statement, and can inform new directions and updates to the framework. Essentially, the evaluation phase of a strategic framework involves asking stakeholders for feedback on progress and adjusting the framework and/or implementation strategies accordingly. This is a cyclical process where actions taken during implementation can have impacts on the institutional context itself, and where feedback can help to inform the next stage of policy development. We suggest the review should take place by consistently examining any gaps developing in research governance, that will inevitably be created by ongoing research innovation and digital disruption. Given the fast pace of change in the capability of generative AI over recent years, it seems inevitable that future changes may necessitate equivalent changes in policy to maintain relevance. We also propose a more direct form of review; to survey stakeholders at regular intervals. 

There are many ways and times that institutions can seek feedback from researchers and other key stakeholders in this review process. In our case, one university surveyed stakeholders views during the development of the position statement with the plan of conducting a thorough survey one year later to compare stakeholder perceptions before and after policy implementation. The other university intended to run a single survey within the first year of implementation. By acting on any identified gaps and improvements from this process, both universities identified that the development of a definitive position statement may be a somewhat transient measure where refinement, redirection, and re-implementation requires ongoing adaptation. Nevertheless, a clear position statement and framework identifying relevant key stakeholders and essential institutional elements within the research context provides critical guidance and structure for institutional governance and policy decision-making.

\section{Conclusion}

As universities grapple with the challenges and opportunities presented by generative AI, our paper introduces a strategic framework to promote responsible and ethical practices in academic research. The generative AI landscape will continue to rapidly evolve, underpinning the urgent need for universities to develop comprehensive and university-wide policy and strategies to ensure research integrity and to adequately support researchers to effectively navigate the responsible use of generative AI.

The framework set out in this paper provides a structured approach, as well as insights into the inter-relating elements that institutions should consider within a practical implementation plan. The central element in this process is the development of a position statement on generative AI, where the key principles of research integrity can be reinforced. This position statement serves as a focus for the operationalisation of training programs, communication strategies, and infrastructure initiatives. The framework outlines a comprehensive iterative process which begins with an examination of contextual factors, prompts the development of the position statement, outlines how this can be implemented, and recommends ongoing review and improvements. 

The framework we offer will, we hope, save other institutions time and costs in strategy development. Readily accessible generative AI has appeared quickly, with its widespread use by researchers potentially preceding current guidelines for how to responsibly do so. Our framework provides much of the preparatory background and institutional structure required to develop policies, procedures, and strategies that are urgently required in this area. As tested by two large universities, the framework provides practical steps for effective implementation and is sufficiently adaptable to suit a range of institutional contexts.

While other practitioners have produced frameworks for dealing with generative AI in universities \citep{luo_critical_2024} broadly, this paper specifically addresses the research context. The framework is a road map for research institutions to utilise and adapt, noting that each institution would need to consider the process within its own particular context to develop strategies and plans to suit its own objectives. Two examples, QUT's and MQ's, of institutional strategies and position statements are provided in this paper to exemplify what the outcomes of a strategic framework process can be, rather than to offer definitive end-points. 

Testing the success of various strategies within this framework could, in fact, form the basis of future research in this area. Future studies could help determine which types of position statements are most effective under certain circumstances, and which strategies best promote responsibility and research integrity in the use of generative AI across different institutions. It will ultimately be up to each organisation to drive its generative AI agenda, but the framework we present in this paper can help initiate and direct that process.

\section*{Supplemental materials}

Attachment 1: Macquarie University's Position statement, available CC-BY at \url{https://zenodo.org/records/10851623}

Attachment 2: Queensland University of Technology's Position Statement, available CC-BY at
\url{https://zenodo.org/records/10971910}

Attachment 3: Chat logs of editing this paper using ChatGPT GPT-4 and Claude 3 Opus, as required by our position statements. (Available at \url{https://osf.io/8dxj6/}. The author order randomisation log is also in this repository.)

\subsection*{Acknowledgements}

We'd like to thank the following early readers for their feedback and encouragement. All errors remain the responsibility of the authors. Our thanks goes to: Jane Thogersen, Stephanie Bradbury, Matt Bower, Sylvie Saab, and Jacqueline Phillips.

\section*{Declarations}

\begin{itemize}
\item Funding

Not applicable.

\item Competing interests 

MH is also a Council Member of the Committee on Publication Ethics and the Director of Tricky Goose Training.

\item Ethics approval 

Not applicable.

\item Consent to participate

Not applicable.

\item Consent for publication

All authors consent to publication. Both universities have decided to make their policies CC-BY with a clear DOI, in addition to their public institutional publication of the policies.

\item Availability of data and materials

All prompts and full chat-logs of sessions with large language models used as part of editing this work are available at https://osf.io/8dxj6/

\item Code availability 

Not applicable.

\item Authors' contributions

SS: Conceptualisation, Writing – Original Draft Preparation and Review and Editing; MT: Conceptualisation, Writing – Original Draft Preparation and Review and Editing; KF: Conceptualisation, Writing – Original Draft Preparation and Review and Editing; AW: Conceptualisation, Writing – Original Draft Preparation and Review and Editing; BBS: Conceptualisation, Writing – Original Draft Preparation and Review and Editing; MH: Conceptualisation, Writing – Original Draft Preparation and Review and Editing; Visualisation; ML: Conceptualisation, Writing – Original Draft Preparation and Review and Editing

\end{itemize}

\noindent

\bibliography{references}

\end{document}